\documentstyle[sprocl,fps]{article}

\def\Journal#1#2#3#4{{#1} {\bf #2}, #3 (#4)}

\def\NPB{{\em Nucl.\ Phys.} B}

\def\PLB{{\em Phys.\ Lett.}  B}
\def\PRL{\em Phys.\ Rev.\ Lett.}
\def\PRD{{\em Phys.\ Rev.} D}
\def\EPJC{{\em Eur.\ Phys.\ J.} C}
\def\NPPS{{\em Nucl.\ Phys.\ [Proc.\ Suppl.]}}
\def\CPC{{\em Comput.\ Phys.\ Commun.}}
\def\HPA{{\em Helv.\ Phys.\ Acta}}
\def\JHEP{{\em JHEP}}
\def\PTP{{\em Prog.\ Theor.\ Phys.}}

\def\be{\begin{equation}}
\def\ee{\end{equation}}
\def\bea{\begin{eqnarray}}
\def\eea{\end{eqnarray}}
\def\lsi{\raise0.3ex\hbox{$<$\kern-0.75em\raise-1.1ex\hbox{$\sim$}}}
\def\gsi{\raise0.3ex\hbox{$>$\kern-0.75em\raise-1.1ex\hbox{$\sim$}}}
\newcommand{\lsim}{\mathop{\lsi}}
\newcommand{\gsim}{\mathop{\gsi}}

\newcommand{\R}{{\kern+.25em\sf{R}\kern-.78em\sf{I} \kern+.78em\kern-.25em}}

\begin{document}

\title{APPROXIMATE GINSPARG-WILSON FERMIONS FOR QCD}

\author{W. BIETENHOLZ}

\address{NORDITA, Blegdamsvej 17 \\ 
DK-2100 Copenhagen, Denmark \\ E-mail: bietenho@nordita.dk} 

\maketitle\abstracts{Lattice fermions obeying the Ginsparg-Wilson
relation do correctly represent the physical properties related
to chirality. This can be achieved by local fermions, which
involve an infinite number of couplings, however. For practical
purposes, it is useful to first construct approximate Ginsparg-Wilson
fermions within a short range. We report on a successful
construction in QCD at $\beta =6$. The good quality of the approximation
is observed from the spectrum, which is situated close to a 
Ginsparg-Wilson circle. These fermions also provide an excellent
approximation to rotational symmetry and they are promising for
a good scaling, since they arise from the perfect action framework.
Their insertion into the overlap formula renders the Ginsparg-Wilson
relation exact. It leads to an improved overlap fermion with a high 
level of locality. This insertion is statistically on safe grounds 
at $\beta \gsim 5.6$.}

\section{Ginsparg-Wilson fermions}

In a slightly simplified form, the famous Nielsen-Ninomiya theorem \cite{NN}
states that a local lattice fermion without species doublers cannot
be chiral in the sense that the lattice Dirac operator $D$
anti-commutes with $\gamma_{5}$. Locality means here that the couplings
in $D$ decay at least exponentially in the separation between
$\psi$ and $\bar \psi$. 
\footnote{Actually the proof in Ref.\ 
[~\raisebox{-6.3pt}[10pt][6pt]{\Large\cite{NN}}\,] 
still holds for an even weaker form of locality.}
Hence it is an obvious idea to break the chiral symmetry by an irrelevant
term, so that it should be restored in the continuum limit.
The simplest way to do so is to set $\frac{1}{2} \gamma_{5} \{D , \gamma_{5} \}$
equal to some local term of $O(a^{2})$,
such as the Wilson term $\frac{1}{2} \Delta$, where $\Delta$ is a
discretized Laplacian. 
\footnote{Here $a$ is the lattice spacing, but in general we will refer to a
hypercubic lattice of unit spacing in Euclidean space.}
However, this type of chiral symmetry breaking on the lattice is rather violent; 
it causes quite some trouble such as additive mass renormalization,
$O(a)$ scaling artifacts, renormalization of currents, mixing of
matrix elements, etc.

On the other hand, it turned out to be harmless to introduce a non-vanishing
anti-commutator as
\begin{equation}
\frac{1}{2} \gamma_{5} \{ D^{-1} , \gamma_{5} \} = R
\end{equation}
where $R$ is a {\em local} term with $\{ R , \gamma_{5} \} \neq 0$.
The superiority of this relation can be understood intuitively
from the fact that $R$ doesn't shift the poles in $D^{-1}$.
In the form
\begin{equation}
\{ D_{x,y} , \gamma_{5} \} = 2 (D \gamma_{5} R D)_{x,y}
\end{equation}
it is known as the {\em Ginsparg-Wilson relation} (GWR) \cite{GW}.

\subsection{Virtues}

Amazingly, it seems that all physical properties related to chirality
are re-presented correctly by a lattice fermion obeying the GWR
(a GW fermion).
The mass and the vector current (as well as the flavor non-singlet
axial vector current) are not renormalized and weak matrix elements
do not mix \cite{Has}. Moreover, the chiral anomalies \cite{GW,ML,anomal} 
as well as global anomalies \cite{BC}, and the soft pion 
theorems are reproduced correctly \cite{soft}. Even the construction of chiral
gauge theories on the lattice is feasible based on GW fermions \cite{MLcg}.

For the understanding of these properties, it is a key observation
that GW fermions have an exact --- though lattice modified ---
chiral symmetry at finite lattice spacing \cite{ML}.

It is instructive to consider the spectrum of a GW fermion.
For simplicity, we assume $D^{\dagger} = \gamma_{5} D \gamma_{5}$
and $R_{x,y} = \frac{1}{2\mu } \delta_{x,y}$ ($\mu > 0$), hence the
GWR reads $\mu (D+D^{\dagger}) = D^{\dagger}D$.
If we introduce the operator $A = D-\mu$, the GWR simplifies further to
\begin{equation}
A^{\dagger} A = \mu^{2} .
\end{equation}
Therefore we know that the spectrum of a GW Dirac operator
is --- with the above assumptions --- always situated on a circle
in the complex plane, with center and radius $\mu$. This confirms
the absence of additive mass renormalization, and it also rules
out ``exceptional configurations'' \cite{Has}. Moreover, it provides
a well-defined index (since the zero eigenvalues are exact),
and together with the index theorem we obtain a sensible
definition of the topological charge of a lattice configuration.

\subsection{Limitations}

After celebrating the impressive properties of GW fermions,
we now have to address their limitations. One point is that the
GWR guarantees a correct chiral behavior, but it does hardly
imply anything about other properties, which are also
essential for a formulation of lattice fermions, in particular
the scaling behavior.

A second point concerns locality: the relaxation of the condition
$\{ D,\gamma_{5}\} =0$ to the GWR allows the lattice fermion
to be local in the sense that the couplings decay exponentially
with the lattice distances. This is sufficient from a conceptual
point of view, since there is a decay length of a finite
number of lattice spacings, which ensures the right continuum
limit. However, for applications one would like to have even
``ultralocality'', which means that the couplings drop to zero
beyond a finite number of lattice spacings. Unfortunately,
GW fermions cannot have this property, not even in the case
of free fermions \cite{ultra}. For example, if we want to insert
the free Wilson Dirac operator and solve for $R$, then we obtain
a pseudo-GW kernel $R$ which decays as $R_{x,y} \propto
\vert x-y \vert ^{-4}$ in $d=2$, and like $R_{x,y} \propto
\vert x-y \vert ^{-6}$ in $d=4$. This is non-local, and therefore the
Wilson fermion does not obey any GWR, not even in the free case
(in the interacting case this is also clear from the mass renormalization).

Of course, in practice one cannot work with couplings
over infinite distances. In a finite volume with certain
boundary conditions, the GWR --- with these boundary conditions
implemented --- can be solved, but this still requires the
coupling of sites (and links) over all distances in this volume.
%(which is inconvenient).

\subsection{Exact and approximate solutions}

Regarding the first limitation,
there exists a class of lattice actions called ``perfect actions''
which deserves its name by yielding a scaling identical to the
continuum at any lattice spacing. At the same time, perfect actions
solve the GWR \cite{GW}, but unfortunately their construction is
about as difficult as solving directly the model under
consideration, since it requires a functional integration
extrapolated to the continuum.

\begin{figure}[htb]
\vspace{-3mm}
\begin{center}
\hspace{10mm}
\def\fpsangle{0}
\epsfxsize=80mm
\fpsbox{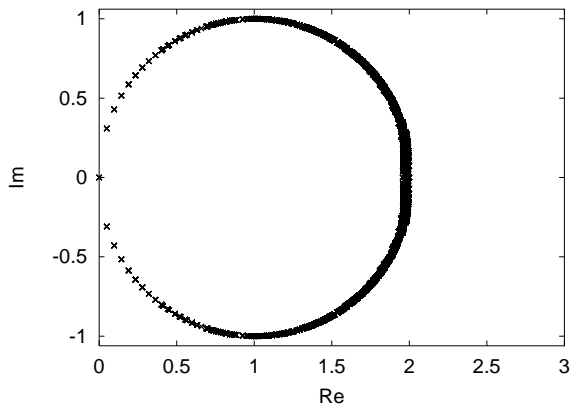}
\end{center}
\vspace{-5mm}
\caption{\it{The spectrum of a truncated perfect, free HF on a $20^4$ lattice
(plotted in $C \!\!\!\! \tiny{I}$).}}
\label{spec4d}
\vspace{-3mm}
\end{figure}

The construction of ``classically perfect actions'' \cite{HN}
is much easier, though still difficult.
They also solve the GWR \cite{Has}, and their
scaling is still excellent. However, a successful construction and
application for interacting fermions could only be achieved in $d=2$
so far \cite{LP}. From the second limitation we know that such actions
need some truncation. Here we truncate to a ``hypercube fermion'',
HF (with couplings not only to nearest neighbor sites, but to all
sites inside a unit hypercube), which is actually applicable in
QCD simulations \cite{HF}. We first consider the truncation for the free
fermion. There the perfect lattice Dirac operator can be constructed,
and the term $R$ occurs in the block variable transformation.
Locality is optimal --- that is, the exponential decay of the couplings is
optimally fast --- for $R_{x,y}= \frac{1}{2} \delta_{x,y}$, \cite{QuaGlu} 
which we denote as the ``standard GW kernel''. Hence the spectrum
of that perfect fermion is situated on a unit circle with center 1.
We now truncate by evaluating the perfect couplings in a small volume 
of $3^{4}$ sites, and then we use the same couplings in any volume 
\cite{BBCW}. We obtain a lattice Dirac operator of the form
\begin{equation}
D(x,y) = \rho_{\mu}(x-y) \gamma_{\mu} + \lambda (x-y) \ ,
\end{equation}
where the support of $\rho_{\mu}(x-y)$, $\lambda (x-y)$ is restricted to
$\vert x_{\nu} - y_{\nu}\vert \leq 1$, ($\nu = 1 \dots 4$).
(These couplings are given in 
Ref.\  [~\raisebox{-6pt}[10pt][6pt]{\Large\cite{BBCW}}\,], Table 1.)
Of course, in a larger volume it is not exactly perfect
any more, and the GWR is violated a little. To probe this truncation effect,
we look at the spectrum of the hypercube fermion on a $20^4$ lattice.
Fig.\ \ref{spec4d} shows that it is indeed very close to a GW circle.
Also the scaling artifacts in this truncated perfect free fermion are 
very small, as we see from the fermion dispersion relation and thermodynamic 
scaling quantities \cite{BBCW,chempot}.

\begin{figure}[htb]
%\vspace{-7mm}
\begin{center}
%\vspace{-5mm}
\hspace{14mm}
\def\fpsangle{270}
\epsfxsize=85mm
\fpsbox{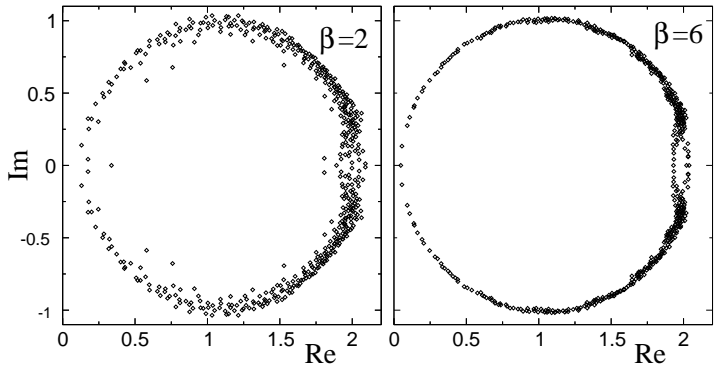}
\end{center}
\vspace{-32mm}
\caption{\it{The spectra of the 2d scaling optimal hypercube fermion (SO-HF)
for typical configurations at strong resp.\ weak coupling,
approximating a GW unit circle.}}
\label{spec2d}
\vspace{-5mm}
\end{figure}

The same construction can be done in $d=2$, but there we prefer to use
a ``scaling-optimal hypercube fermion'' (SO-HF) \cite{WBIH}, which is similar to
the truncated perfect one, but still a bit improved (with respect
to scaling). Again the spectrum for the free fermion is close to
a GW unit circle, and its scaling is excellent.
To simulate this lattice fermion in the 2-flavor Schwinger model, we performed a
``minimal gauging'' by hand: we attached the free couplings to the shortest
lattice paths only, in equal parts where several shortest paths exist.
We also added a clover term with coefficient 1, and used the standard plaquette
gauge action. Of course, this simple gauging brings in additional artifacts and
a further deviation from the GWR. Hence the eigenvalues spread more and more
around the unit circle as $\beta$ decreases, but for instance at $\beta =2$
the circle is still approximated more or less, see Fig.\ \ref{spec2d}.

\begin{figure}[hbt]
\vspace{-2mm}
\begin{tabular}{cc}
\hspace{-4mm}
\def\fpsangle{0}
\epsfxsize=58mm
\fpsbox{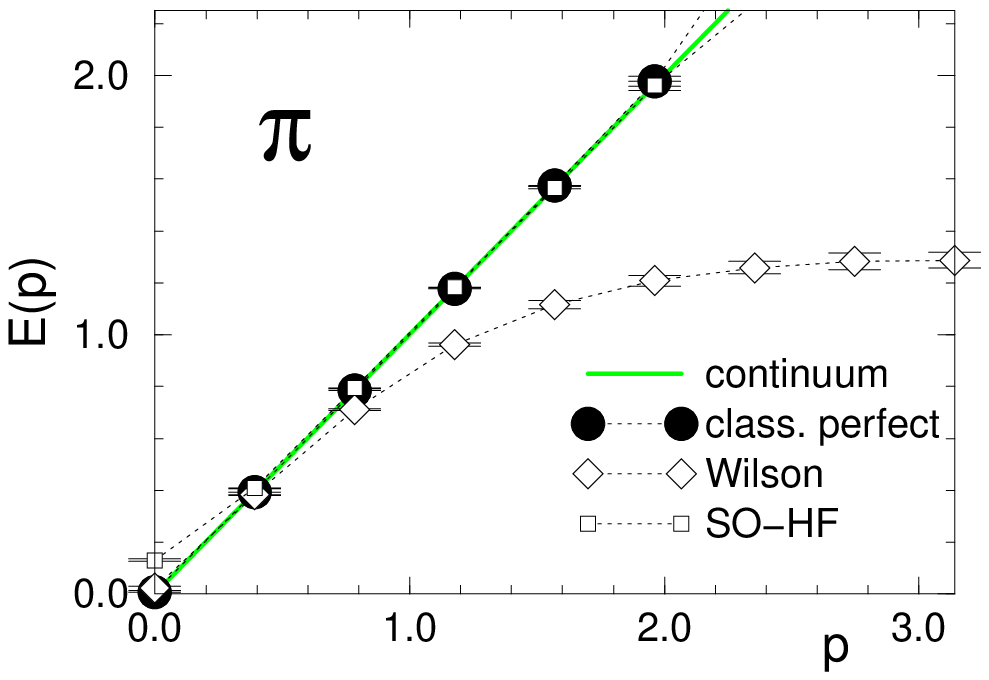}
%\hspace{5mm}
\def\fpsangle{0}
\epsfxsize=58mm
\fpsbox{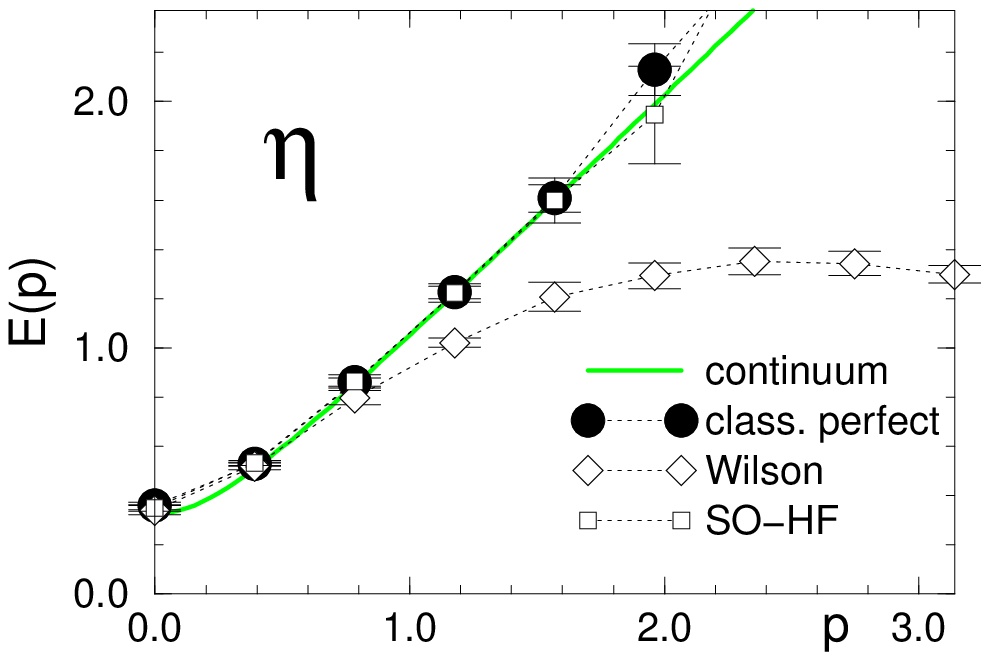}
\end{tabular}
\vspace{-5mm}
\caption{\it{The ``meson'' dispersion relations for three types of
lattice fermions in the 2-flavor Schwinger model at $\beta = 6$.}}
\vspace{-2mm}
\label{meso-disp}
\end{figure}

As a scaling test we measured the dispersion relations of the two meson-type
states, a massless triplet and a massive singlet \cite{SaWi},
which we denote as $\pi$ and $\eta$. Indeed, the scaling is dramatically
improved over the Wilson fermion (at critical hopping parameter), see Fig.\
\ref{meso-disp}. Amazingly, this SO-HF reaches a similar scaling quality
as the classically perfect action, which was truncated only very mildly to
123 independent couplings per site \cite{LP}. In contrast, the SO-HF involves
only 6 independent couplings per site, hence this approach has the
potential to be extended and applied in $d=4$.

\begin{figure}[htb]
\vspace{-5mm}
\begin{center}
\hspace{15mm}
\def\fpsangle{0}
\epsfxsize=60mm
\fpsbox{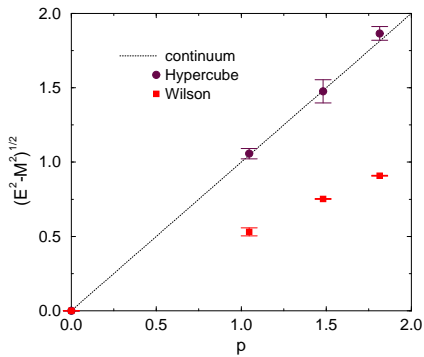}
\end{center}
\vspace{-5mm}
\caption{\it{The ``speed of light'' $c = p/\protect\sqrt{E^2 -M^2}$ ($M =$ ``pion mass'')
in QCD at $\beta=5$. \protect\cite{BBCW}}}
\label{spofl}
\vspace{-6mm}
\end{figure}

Also the rotational symmetry is approximated very well for such HFs.
We tested this for free and for interacting fermions in $d=2$ \
\cite{WBIH} and $d=4$ \ \cite{BBCW}. As an example, we show the ``speed of
light'' for the minimally gauged truncated perfect HF in QCD at $\beta =5$,
and compare it to the Wilson fermion in Fig.\ \ref{spofl}.
What is not visible from that figure, however, is the dramatic
mass renormalization: what is supposed to be the pion mass amounts
to $M=3.0$. Also in the Schwinger model this problem is serious,
as Figs.\ \ref{spec2d} and \ref{meso-disp} show: at $\beta =6$ we 
obtain a $\pi$ mass of 0.13.
This is the one unpleasant feature of the otherwise successful HFs
with minimal gauging.
In the next section we discuss a possibility to eliminate
this effect completely. In Sec.\ 4 we are going to consider further
methods to approach the chiral limit.

\section{Improved overlap fermions}

Let us start from some lattice Dirac operator $D_{0}$, which
obeys the conditions of the Nielsen-Ninomiya theorem (local,
no doublers etc.). We assume again $D_{0}^{\dagger} = \gamma_{5}
D_{0} \gamma_{5}$, and we recall that the GWR for
$R_{x,y} = \frac{1}{2\mu} \delta_{x,y}$ is equivalent to
$A_{0}^{\dagger} A_{0} = \mu^{2}$, for $A_{0}=D_{0}-\mu$, $(\mu >0)$.
In general this will not hold, of course, but we can simply enforce it
by the overlap formula
\begin{equation} \label{overlap}
A_{ov} = \frac{\mu A_{0}}{\sqrt{ A_{0}^{\dagger} A_{0} }} \ ; \quad
D_{ov} = A_{ov} + \mu \ .
\end{equation}
$D_{ov}$ represents a GW fermion. A prototype was proposed by
H.\ Neuberger \cite{Neu}, starting from the Wilson fermion
$D_{0}=D_{W}$ and $\mu =1$,
\begin{equation}
D_{Ne} = 1 + \frac{A_{W}}{\sqrt{ A_{W}^{\dagger} A_{W} }} \ ; \quad
A_{W} = D_{W} -1 \ .
\end{equation}
In the free case and in a smooth gauge background, $D_{Ne}$ is
free of doublers and local. The latter has been established analytically
assuming a constraint on each plaquette variable \cite{HJL,Neu-loc}, 
and numerically for QCD at $\beta =6$ \ \cite{HJL}. For first simulations
in quenched QCD, see 
Refs.\ [~\raisebox{-6pt}[10pt][6pt]{\Large\cite{simu,chin}}\,].

However, this is just one example in a large class of overlap fermions,
which is obtained by varying $D_{0}$ and $\mu$. \cite{EPJC}
In particular, if it happens that $D_{0}$ represents a GW fermions already,
then it is just reproduced under the overlap formula, $D_{ov}=D_{0}$
(for a fixed GW kernel $R$). Therefore, also perfect and classically perfect
fermions are special types of overlap fermions.
In practice we do not have an exact GW operator $D_{0}$ at hand,
but we can construct an approximation, as discussed in Secs.\ 1.3 and 4.
Then the square root will keep close to the constant $\mu$, and 
$D_{ov} \approx D_{0}$. We may say that the overlap formula provides
a ``GW correction'' of $D_{0}$; in particular it removes the additive 
mass renormalization.

Based on this property, we suggest the following concept \cite{EPJC}:
{\em start from a truncated perfect fermion (or something similar), which
scales well and approximates the GWR, and insert it as $D_{0}$ into the 
overlap formula}. The resulting $D_{ov}$ has exact GW chirality (in particular
we are in the chiral limit) {\em and} a good scaling behavior can also be
expected, due to $D_{ov} \approx D_{0}$. The latter also suggests that
an approximate rotational symmetry of $D_{0}$ is essentially inherited by
$D_{ov}$, and that $D_{ov}$ is very local in the sense of a fast exponential
decay: remember that $D_{0}$ is ultralocal, and --- due to the modest modification
--- the long-range couplings will be turned on just a little bit in $D_{ov}$.
\footnote{Also $D_{W}$ is ultralocal, but not an approximate GW fermion,
so its change to $D_{Ne}$ is rather drastic
and the argument for a high level of locality does not apply to that case.}

To summarize, this concept aims at combining all desirable properties of
a lattice fermion formulation. It has been tested comprehensively in the
two-flavor Schwinger model, and the above expectations are
observed to hold in a very satisfactory way. In rest of this Section
we summarize our 2d results from 
Ref.\  [~\raisebox{-6pt}[10pt][6pt]{\Large\cite{WBIH}}\,]. 

\begin{figure}[hbt]
\vspace{-1mm}
\begin{tabular}{cc}
\hspace{-4mm}
\def\fpsangle{0}
\epsfxsize=58mm
\fpsbox{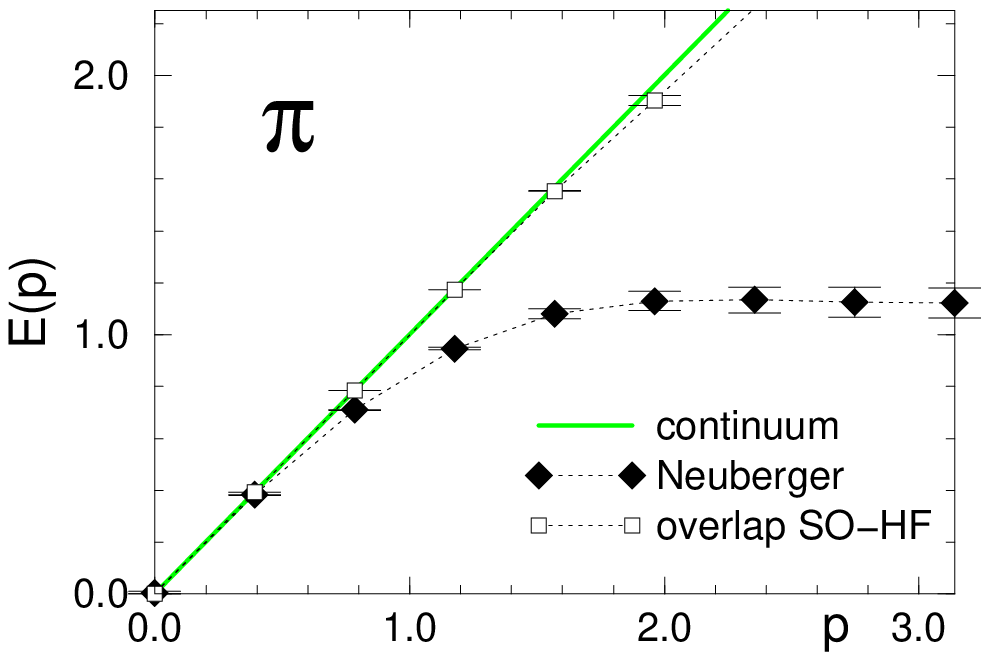}
%\hspace{5mm}
\def\fpsangle{0}
\epsfxsize=58mm
\fpsbox{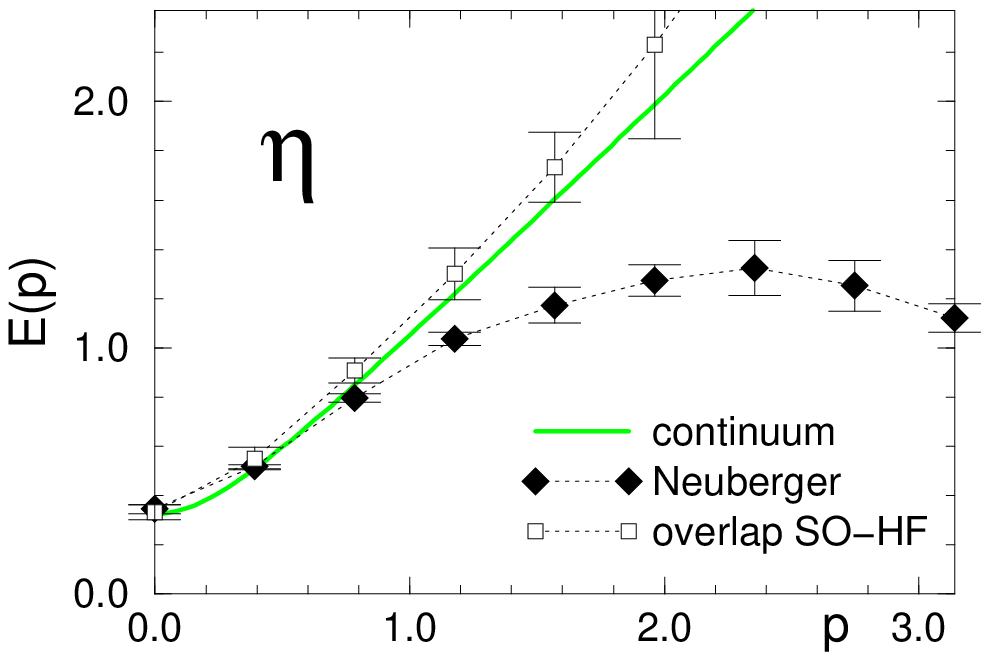}
\end{tabular}
\vspace{-5mm}
\caption{\it{The ``meson'' dispersion relations for different types
of overlap fermions in the 2-flavor Schwinger model at $\beta = 6$.}}
\vspace{-3mm}
\label{meso-ov-disp}
\end{figure}

We first checked the scaling quality of the free overlap fermions
(fermionic dispersion, thermodynamic scaling ratios)
and it turns out that the improvement of the SO-HF over the Wilson
fermion persists under the overlap formula, i.e.\ the overlap SO-HF
is strongly improved over the Neuberger fermion. By comparing the
spectrum of a fixed configuration for the SO-HF and for the overlap
SO-HF, we could then literally see that the alteration due to the
overlap formula is small; the eigenvalues are moved almost radially
onto the unit circle. Next we tested the scaling in the interacting
case, and  the drastic improvement of the overlap SO-HF over the
Neuberger fermion is confirmed again, see Fig.\ \ref{meso-ov-disp}.
We have used standard operators in both cases, hence the comparison
is fair. Of course, both types of overlap fermions may improve if
one consistently improves the operators, which can be tedious, however.
Then one expects also the Neuberger fermion to scale better
\cite{Capi,chin} (in particular $O(a)$ artifacts are excluded 
\cite{Ferenc}), but our results imply that improved overlap fermions 
can be used very successfully even with the simple standard operators
(as it was also observed for the classically perfect action \cite{HN,LP}).

\begin{figure}[hbt]
%\vspace{-3mm}
\begin{tabular}{cc}
%\vspace{-3mm}
\hspace{-5mm}
\def\fpsangle{270}
\epsfxsize=41mm
\fpsbox{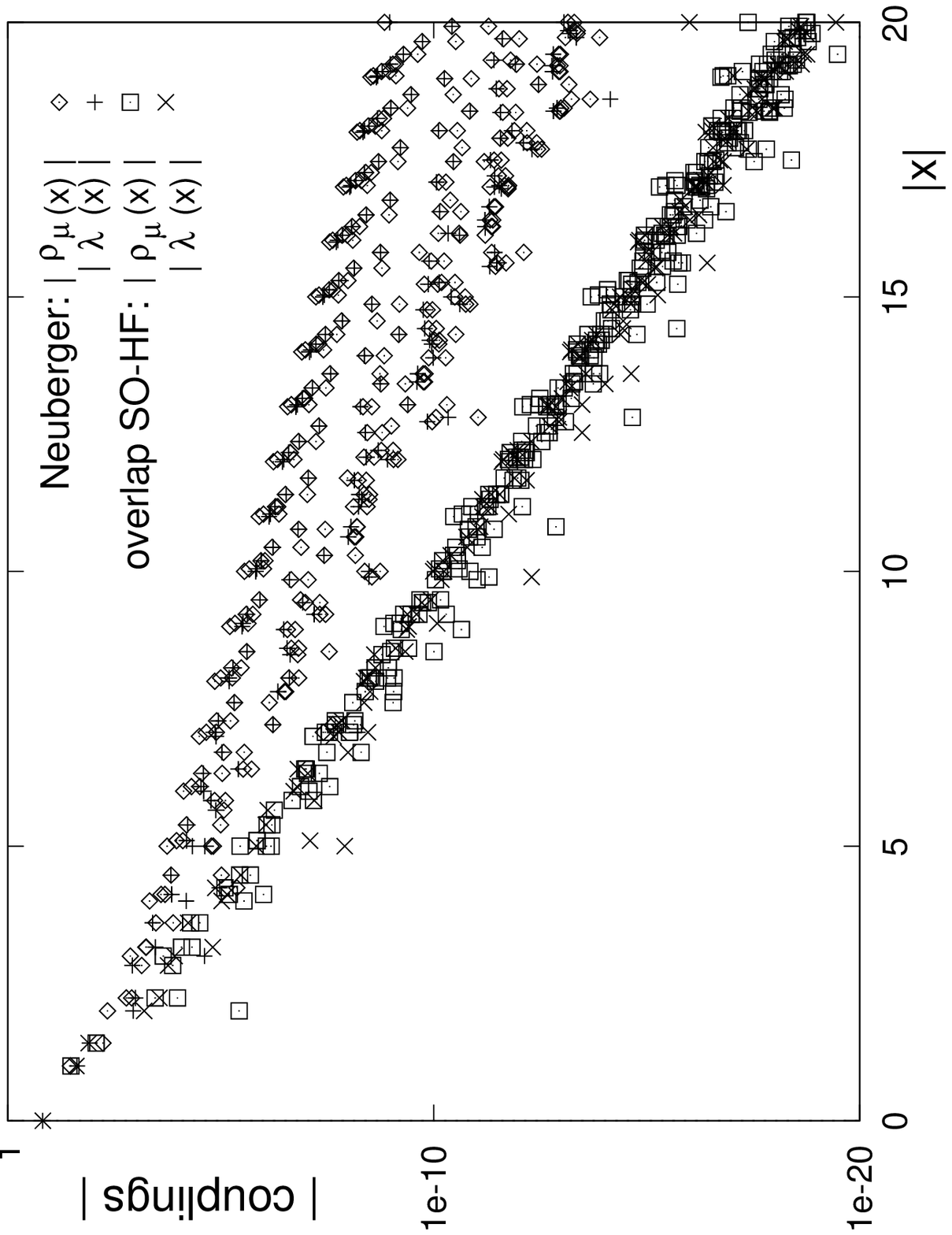}
%\hspace{5mm}
\def\fpsangle{0}
\epsfxsize=65mm
\fpsbox{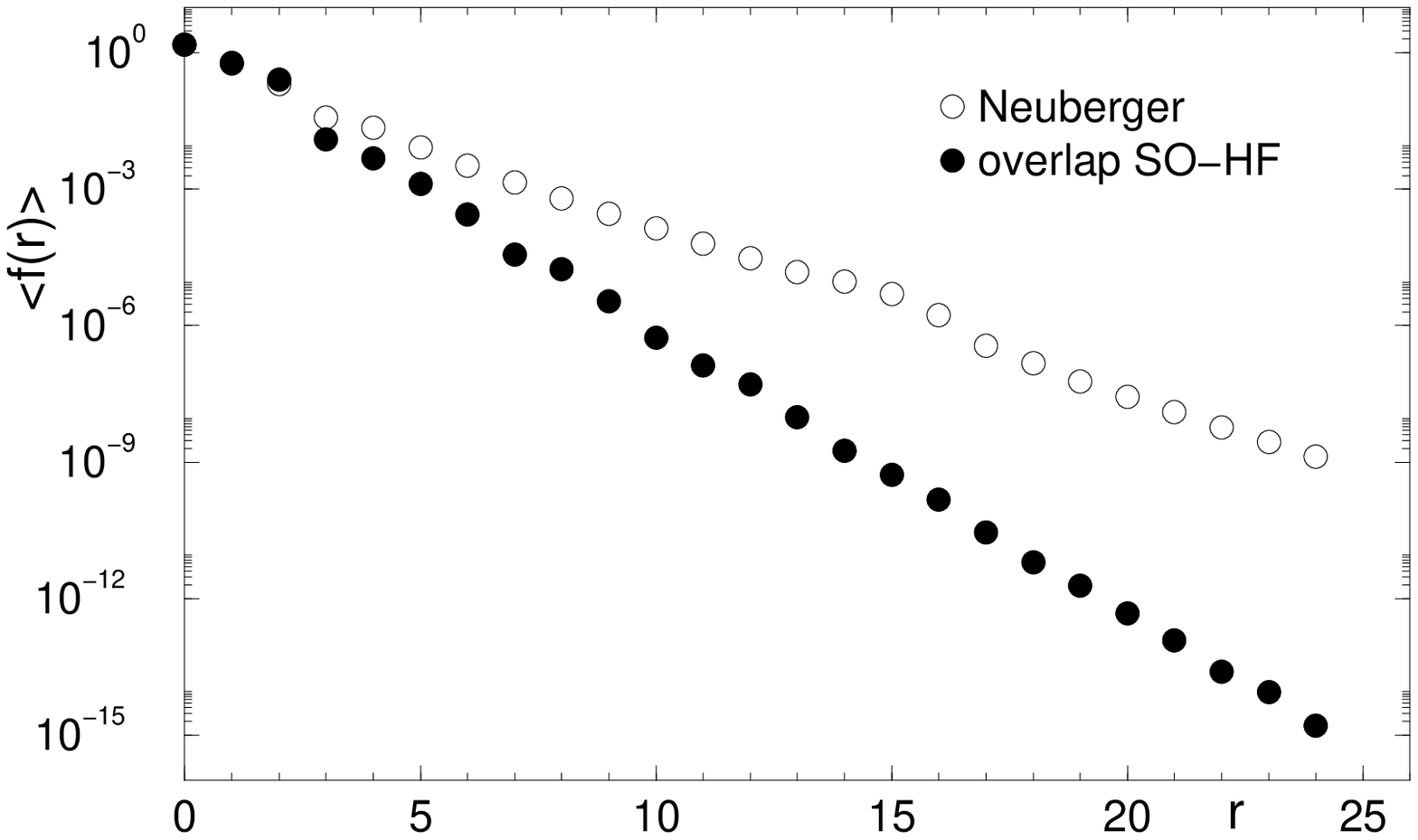}
\end{tabular}
\vspace{-4mm}
\caption{\it{The level of locality for the Neuberger fermion compared to
the overlap SO-HF: the decay of the free couplings in all directions
(left) and the decay of the ``maximal correlation'' \protect\cite{HJL}
over a distance $r$ at $\beta =6$ (right).
The width of the ``cones'' in the left figure also shows how well
rotation invariance is approximated.}}
\vspace{-3mm}
\label{loc-fig}
\end{figure}

Finally we also tested our prediction regarding the degree of locality,
and we found that indeed the exponential decay is much faster for the
overlap SO-HF than for the Neuberger fermion, see Fig.\ \ref{loc-fig}
and Table \ref{loc-tab}. 
\footnote{Ref.\ [~\raisebox{-6pt}[10pt][6pt]{\Large\cite{Capi}}\,] 
suggests to use
non-standard operators (with $\langle R\rangle $ being subtracted 
from $\langle D^{-1}\rangle $)
even for the free fermion, which does improve e.g.\ the scaling
of the free Neuberger fermion. 
However, one then deals with a non-local fermion, even though it
is used only in an indirect way, which raises conceptual questions
concerning the continuum limit. An exception is the perfect fermion at 
$R=0$, where for instance the axial anomaly is reproduced correctly \cite{BWPLB}.}

\begin{table}
\begin{center}
\begin{tabular}{|c|c|c|c|c|c|}
\hline
& & $\!\!\!\!\!\!$ & perfect & overlap-HF & Neuberger \\
\hline
\hline
$d=2$ & $r_{\rho}$ & $\!\!\!\!\!\!$ & 1.268 & 1.255 & 1.930 \\
\hline
$d=2$ & $r_{\lambda}$ & $\!\!\!\!\!\!$ & 0.871 & 0.888 & 1.248 \\
\hline
\hline
$d=4$ & $r_{\rho}$ & $\!\!\!\!\!\!$ & 1.635 & 1.519 & 2.530 \\
\hline
$d=4$ & $r_{\lambda}$ & $\!\!\!\!\!\!$ & 1.187 & 1.109 & 1.708 \\
\hline
\end{tabular}
\end{center}
\caption{\it{The characteristic radius
$r_{\lambda} = ( \sum_{x} \vert \lambda (x) \vert x^{2})/
( \sum_{x} \vert \lambda (x) \vert)$ and (analogously) $r_{\rho}$ 
for various types of free GW fermions.}}
\vspace{-4mm}
\label{loc-tab}
\end{table}

\section{Overlap fermions and the doubling problem}

As we mentioned in Sec.\ 2, overlap fermions are free of doublers
at least in a smooth gauge background. It should be clarified, however,
that the overlap formula itself does not remove any doublers.
They have to be removed before, in $D_{0}$, by something
like a Wilson term; in our ansatz $D_{0}=\rho_{\mu}\gamma_{\mu} + \lambda$,
the scalar term $\lambda$ is crucial for that purpose. Then the overlap formula 
cures the chiral symmetry again in the sense of the GWR, while doublers are 
supposed not to be re-introduced.

We keep on referring to a point-like GW kernel $R_{x,y}=\frac{1}{2\mu} \delta_{x,y}$,
where the mass parameter $\mu >0$ can be chosen. It is the center of the circle
through zero that the spectrum of the Dirac operator is mapped on by the
overlap formula (which also involves the parameter $\mu$).
Let us focus on the eigenvalues of $D_{0}$ close to the real axis: 
the small (almost) real eigenvalues
have to be mapped on (the vicinity of) zero, whereas the eigenvalues with
large real parts have to be mapped on the opposite arc, i.e.\
on (the vicinity of) $2 \mu$. For weak up to moderate coupling,
there is a (statistically) safe interval where $\mu$ can be chosen such
that this separation is achieved. Inside this interval, $\mu$ may be optimized
with respect to various criteria; the optimal $\mu$ tends to rise with increasing
coupling strength --- since it has to adapt to the mass renormalization ---
as has been observed by optimizing locality \cite{HJL} or minimizing
the mapping effect \cite{WBIH}. For instance, in the Schwinger model
such a safe interval still exists at $\beta =2$, as Fig.\ \ref{spec2d}
illustrates.

\begin{figure}
%\vspace{-3mm}
\begin{tabular}{cc}
\hspace{-8mm}
\def\fpsangle{270}
\epsfxsize=60mm
\fpsbox{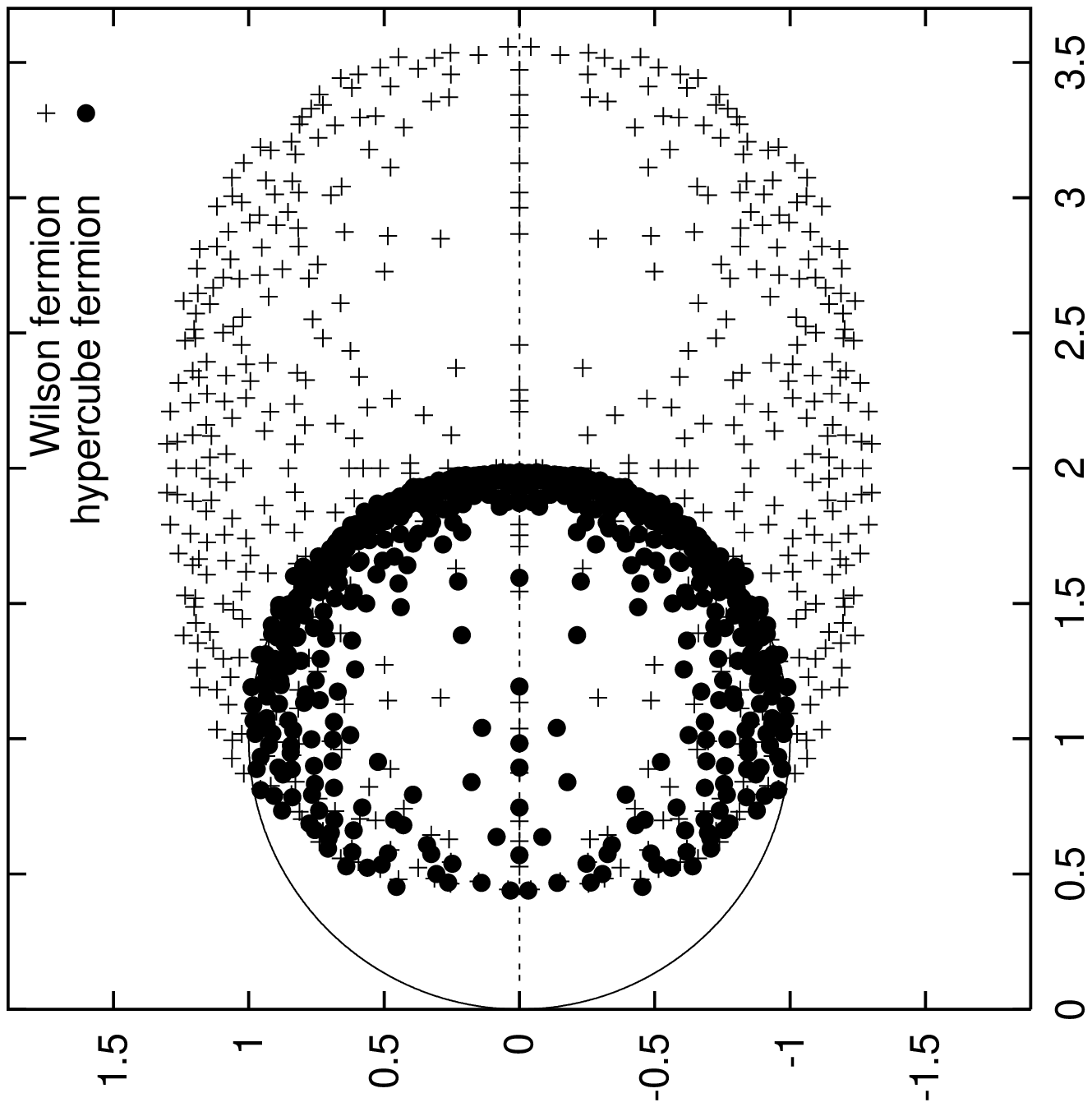}
\hspace{-5mm}
\def\fpsangle{270}
\epsfxsize=60mm
\fpsbox{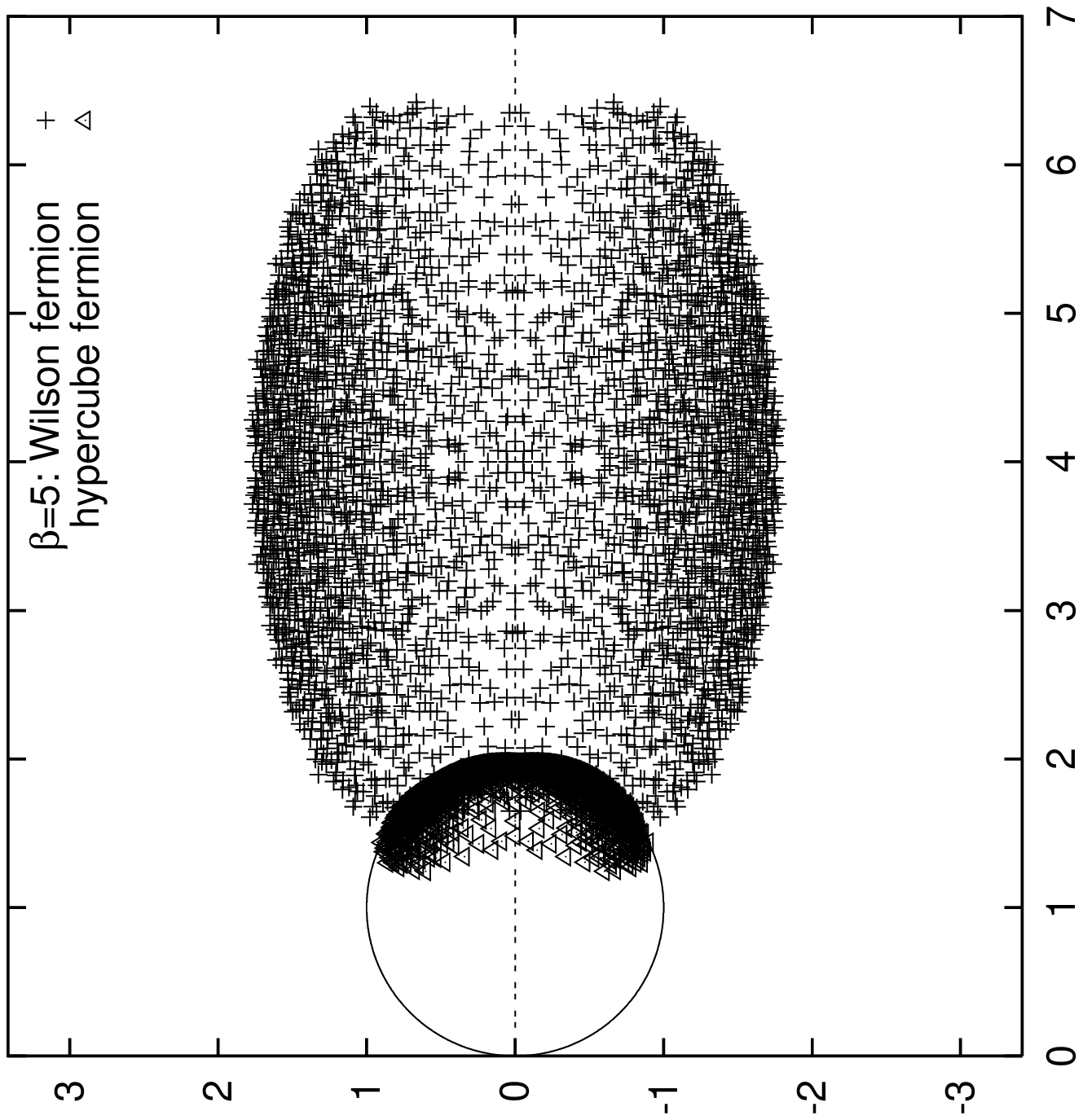}
\end{tabular}
\vspace{-4mm}
\caption{\it{Typical spectra of minimally gauged HFs and 
Wilson fermions (with the hopping parameter
of the free fermion) in the Schwinger model at $\beta =1$ (left) 
and in QCD at $\beta =5$ (right). This illustrates situations
where the overlap projection is not applicable any more.}}
\vspace{-5mm}
\label{strong-fig}
\end{figure}

But how about really strong coupling ? At some point, the eigenvalues 
of typical configurations spread all over inside a certain area for
simple (short-ranged) operators $D_{0}$. This is the case for the Wilson fermion
and for the HF in the Schwinger model at $\beta =1$, and for QCD at $\beta=5$, 
\cite{Dubna} as Fig.\ \ref{strong-fig} shows.
%\footnote{The author had to work hard to get this preprint number $\dots$}.
Wherever we choose $\mu$, mappings to the ``wrong'' arc will occur
frequently; the possibility of ``wrong'' projections
implies that both, the doubling problem (too many projections to the left arc)
as well as mass renormalization (too many projections to the right arc)
are back. Hence at really strong coupling, both of these
problems return for the overlap fermions based on some simple $D_{0}$,
as was also revealed by a strong coupling expansion \cite{strong}
\footnote{The question if also locality breaks down at some point
of relatively strong coupling, where the overlap formula is
still applicable, is currently under investigation \cite{prep}. 
For instance, one can check this for the Neuberger fermion in the 
Schwinger model at $\beta \lsim 2$, or in QCD at $\beta \approx 5.6$.}.
Probably the only way out of this problem would be an excellent approximation
to a (classically) perfect fermion, which is, however, very difficult
to construct and implement.

\begin{figure}[hbt]
%\vspace{-3mm}
\begin{tabular}{cc}
\hspace{-8mm}
\def\fpsangle{270}
\epsfxsize=60mm
\fpsbox{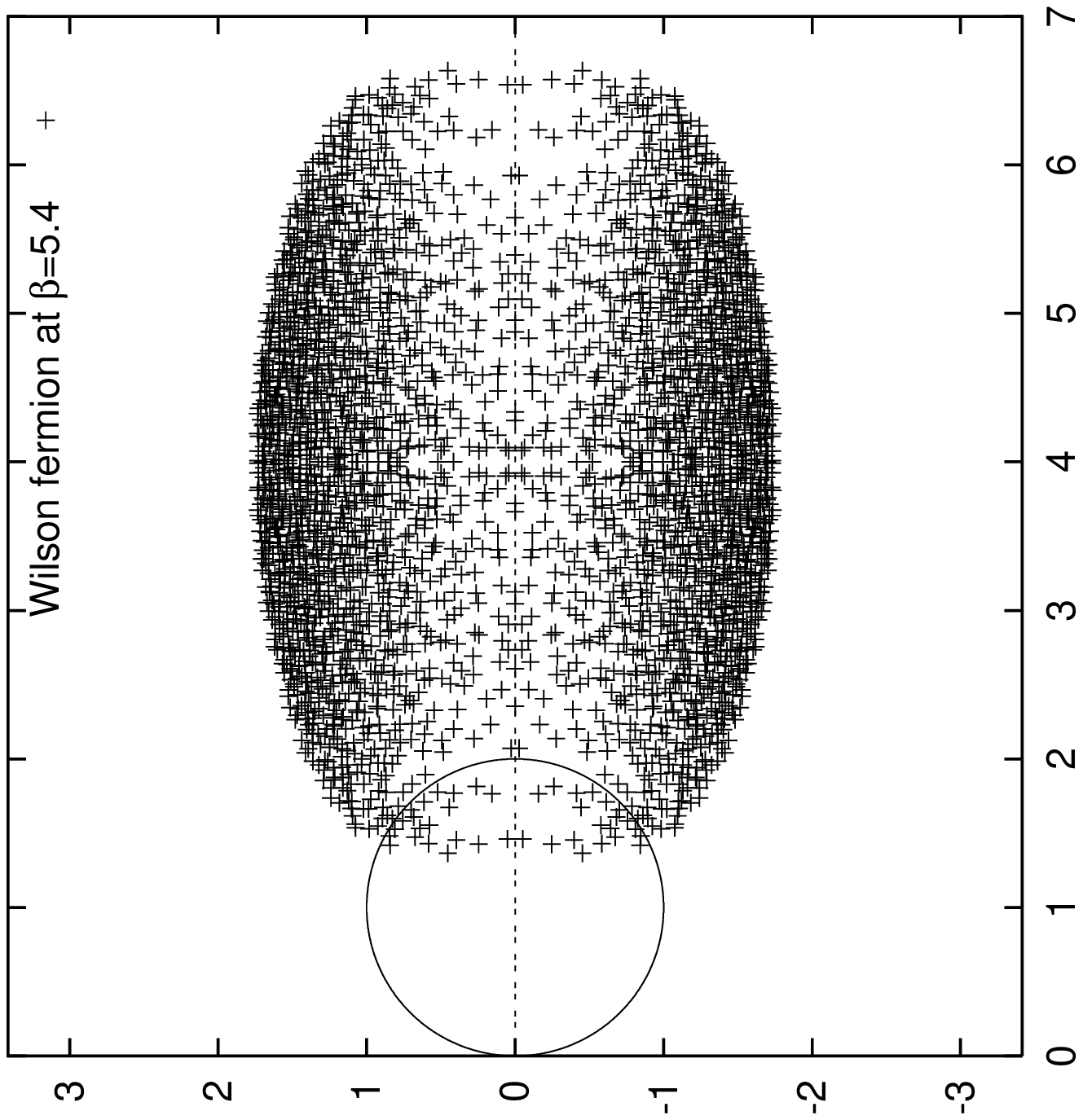}
\hspace{-5mm}
\def\fpsangle{270}
\epsfxsize=60mm
\fpsbox{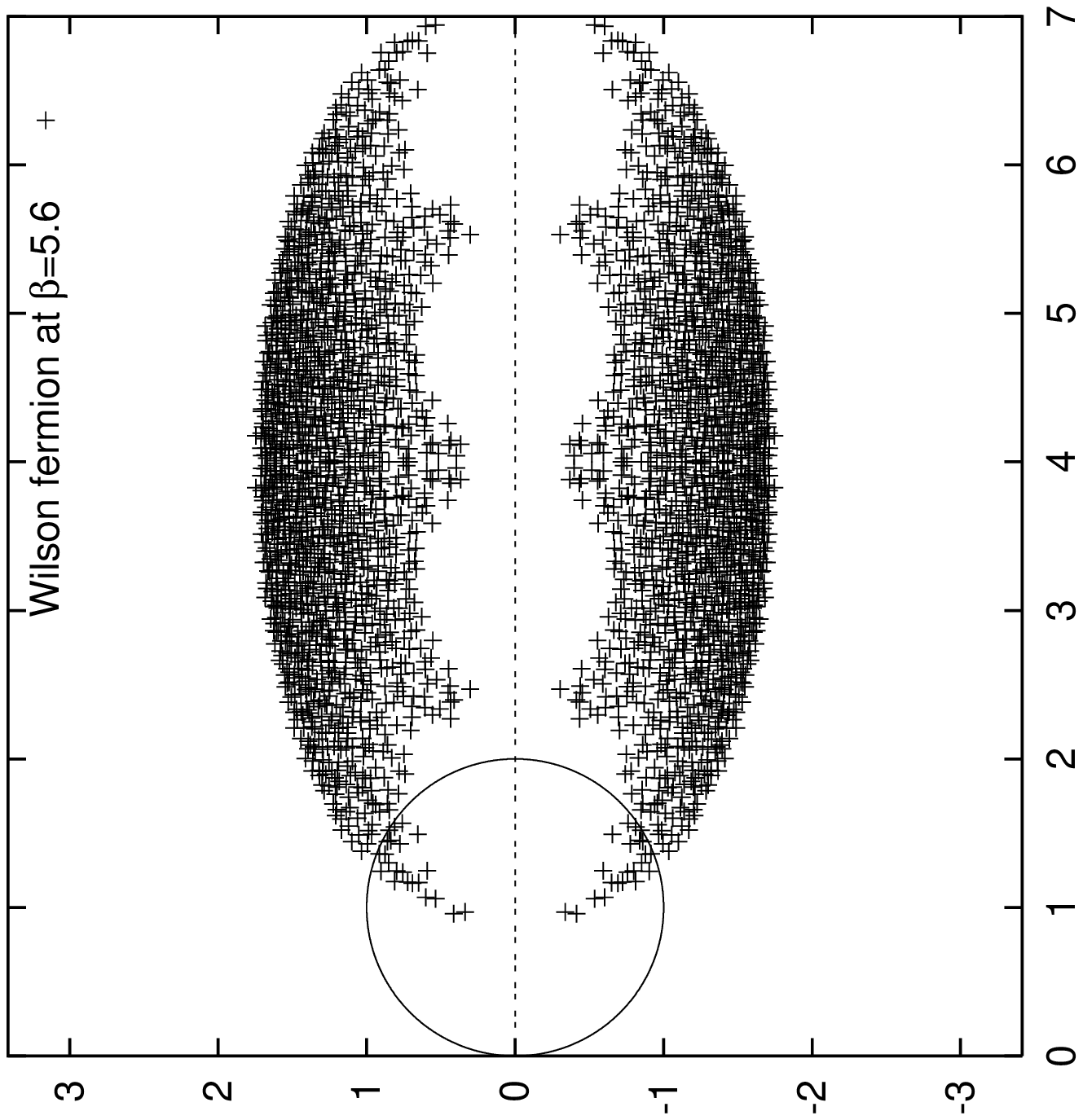}
\end{tabular}
\vspace{-4mm}
\caption{\it{Typical spectra of the Wilson fermion in QCD on a $4^{4}$ lattice
at $\beta =5.4$ (left) and $\beta =5.6$ (right). We recognize the transition
to the regime where the overlap projection is statistically safe 
(for $1.5 \lsim \mu \lsim 2$).}}
\vspace{-4mm}
\label{beta5.4-5.6}
\end{figure}

\begin{figure}[hbt]
%\vspace{-3mm}
%\begin{center}
\hspace{28mm}
\def\fpsangle{270}
\epsfxsize=50mm
\fpsbox{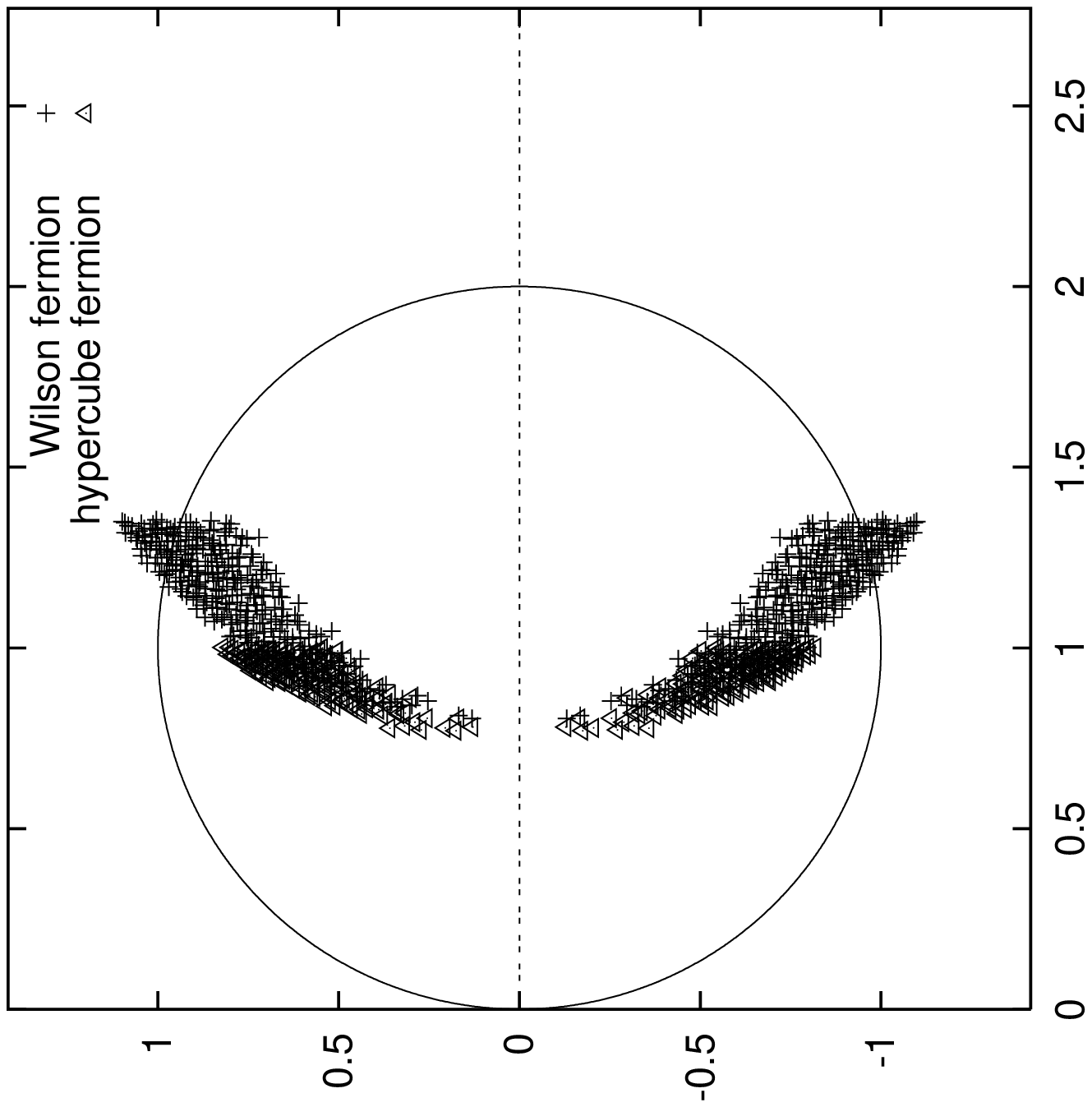}
%\end{center}
\vspace{-3mm}
\caption{\it{The left arc of typical spectra in QCD at $\beta =6$ on an $8^{4}$
lattice for the Wilson fermion and the minimally gauged HF.
We show the 600 resp.\ 300 eigenvalues with smallest real parts, and we
see that here the left arc is manifestly isolated.}}
\vspace{-4mm}
\label{leftarc}
\end{figure}

It is now of interest to get an idea where this transition occurs.
Fig.\ \ref{beta5.4-5.6} shows the Wilson spectra for typical (quenched)
QCD configurations at $\beta =5.4$ and $\beta = 5.6$. We see that the
existence of a safe interval for $\mu$ seems to set in in between.
\footnote{From our experience, even lattices as small as $4^{4}$ provide
a reliable insight into such questions as the applicability of the overlap
formula and the level of approximation to the GW circle. A disadvantage
is that the left arc close to zero is missing --- not due to the lattice action 
but just due to small volume. Therefore we also evaluated this subset of the 
spectra (the eigenvalues with smallest real parts) on $8^{4}$ lattices, see below.}
For the HF the situation is similar, just a little better. Hence $\beta =6$
can be regarded a really safe regime with respect to this issue ---
Fig.\ \ref{leftarc} shows that an isolated left arc is present --- and also
with respect to locality \cite{HJL}, as we mentioned before.

We therefore concentrate on QCD at $\beta =6$ in the next Section, 
and we show that we can construct HFs, which approximate the GWR well
at that coupling strength. This is the crucial ingredient needed
to carry on the program of improved overlap fermions --- 
described in Sec.\ 3 --- to QCD. 

Note that in $d=4$ the notorious square
root has to be evaluated by some iterative method, and starting from a
good GW fermion approximation instead of the Wilson fermion is also
highly profitable with that respect: starting in the right vicinity
speeds up the convergence a lot, as we see for instance from the perturbative
expansion of the square root around $\mu$. \cite{WBIH}
%(In the transition from the Wilson to the Neuberger fermion,
%one risks to perform most iteration steps in a region far from the GWR.)

\section{Approximate Ginsparg-Wilson Fermions for QCD}

Our approach to construct a short-ranged approximate GW fermion for QCD
is to stay with the couplings of the truncated perfect free fermion and
gauge it by hand, using just very few new parameters to go beyond the
``minimal gauging''.
This concept was successful in $d=2$, and in $d=4$ we already
know that the free HF is doing well in scaling, approximating the GWR
and approximating rotational invariance (the latter is also checked at strong
coupling), see Sec.\ 1.3. Alternatively, one may try to minimize the GWR 
violation directly within a limited set of parameters \cite{WBIH,HG},
or undertake a new effort to parameterize an (approximate) classically 
perfect action \cite{Bern}. 

We are confident that our free HF couplings already provide a good
scaling, so the issue is to find a suitable gauging in the sense that
the GWR violation is small. As our criterion, we compute the spectra on
small lattices and try to arrange for them to be close to a GW unit 
circle (for typical quenched configurations at $\beta =6$).

\begin{figure}
%\vspace{-3mm}
\begin{tabular}{cc}
\hspace{-8mm}
\def\fpsangle{270}
\epsfxsize=60mm
\fpsbox{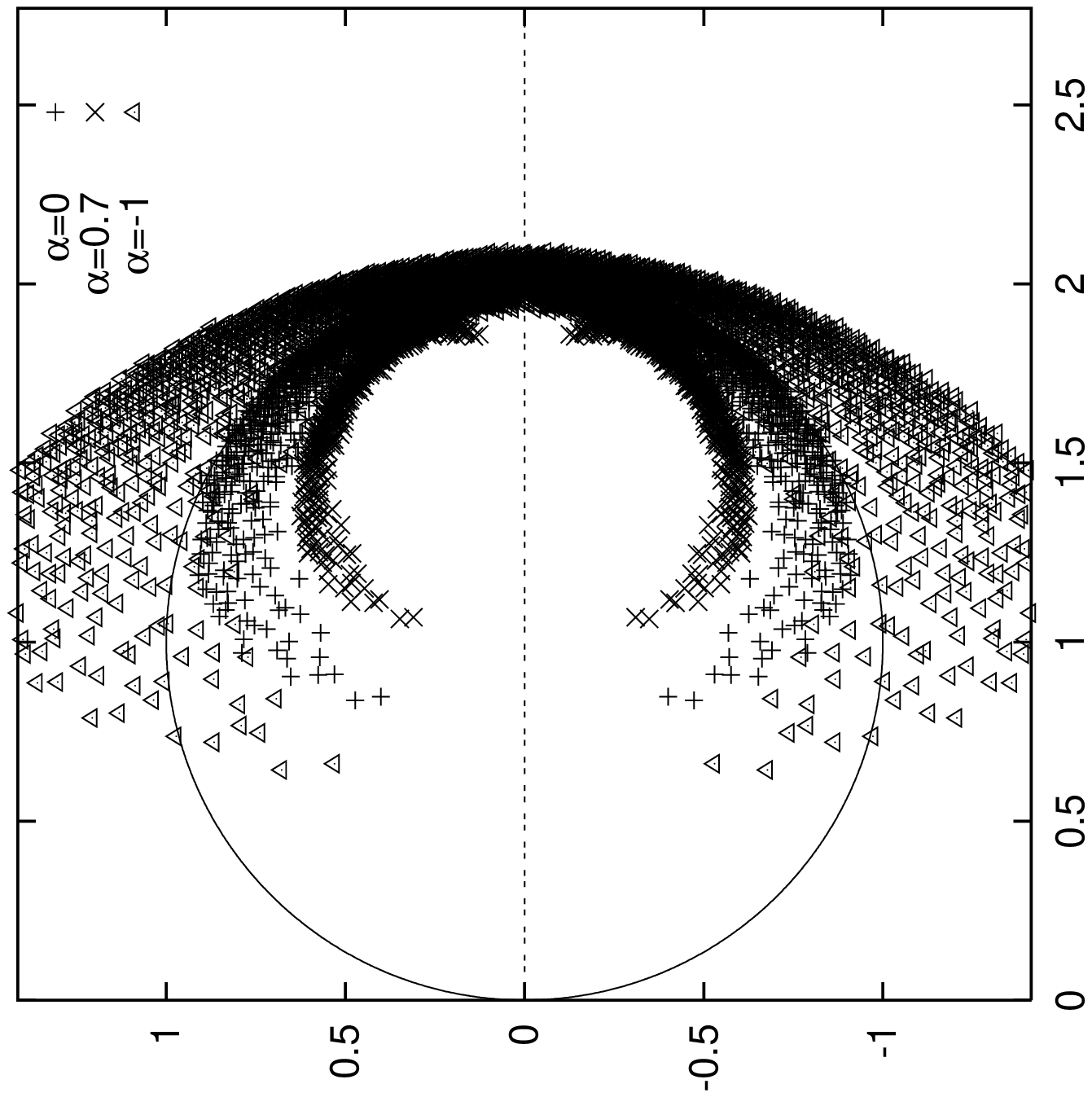}
\hspace{-5mm}
\def\fpsangle{270}
\epsfxsize=60mm
\fpsbox{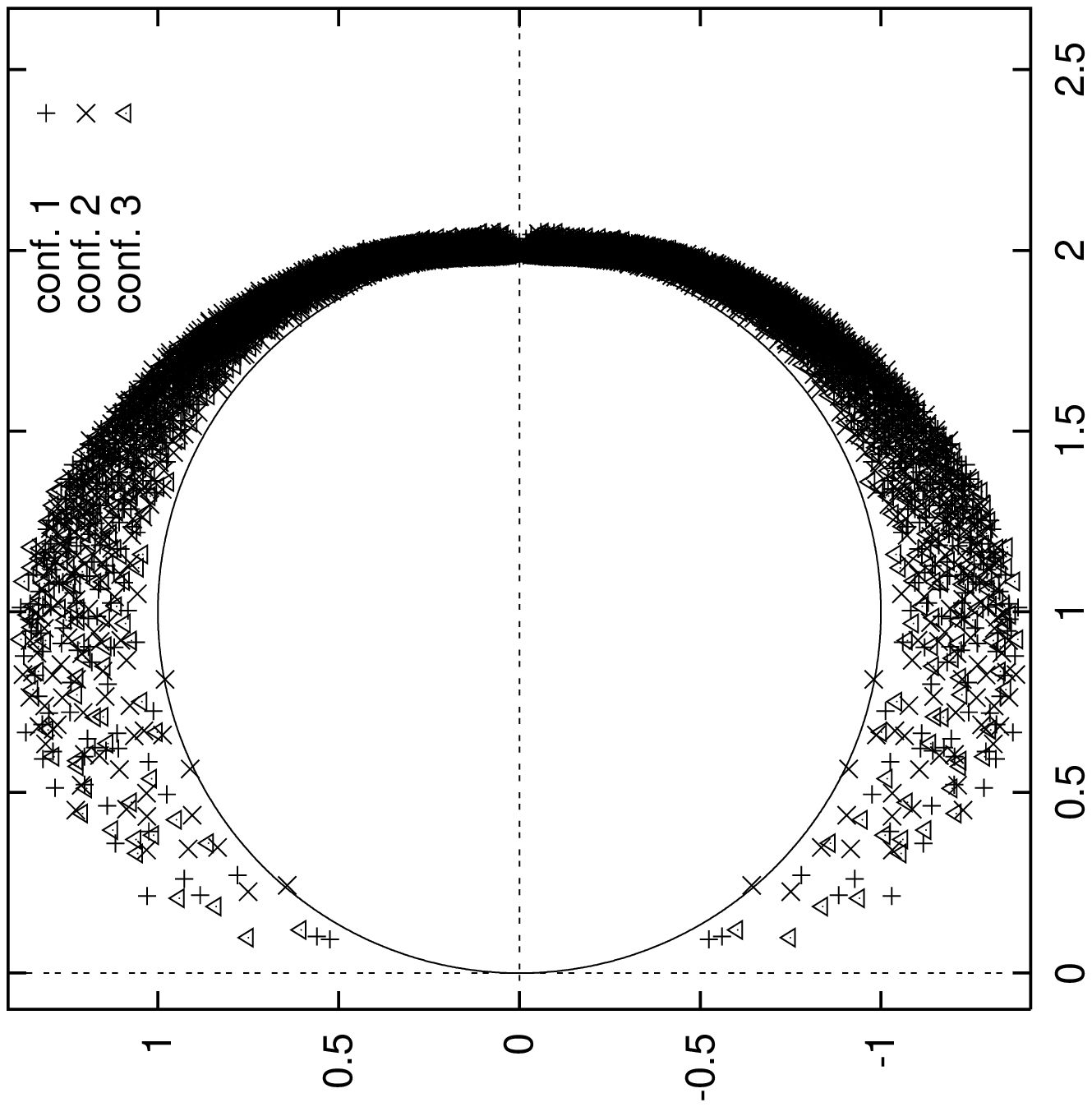}
\end{tabular}
\vspace{-3mm}
\caption{\it{The effect of fat links (left, $u=1$) and link amplification
(right, $\alpha =0, u=0.8$) for a HF spectrum in QCD at $\beta =6$ on a 
$4^{4}$ lattice.}}
\vspace{-5mm}
\label{stap-ucrit}
\end{figure}

As we see from Fig.\ \ref{leftarc}, the minimally gauged HF
suffers from mass renormalization almost as much as the Wilson
fermion. On the other hand, the right arc is excellent
(see Fig.\ \ref{strong-fig}), but less important.
Our first step beyond minimal gauging is the use of fat links:
each link in a given configurations is substituted as
\begin{equation}
link ~ \to ~ (1-\alpha ) ~ link + \frac{\alpha}{6} ~ [ ~ \sum ~ staples ~ ] 
\qquad (\alpha \in \R ) .
\end{equation}
This is computationally cheap, and since we perform just one such
fattening step we do not need to project back onto the gauge group
(in contrast to Ref.\ 
[~\raisebox{-6pt}[10pt][6pt]{\Large\protect\cite{TDG}}\,]).
The results for different $\alpha$ are shown in Fig.\ \ref{stap-ucrit} (left).
As we observed already in $d=2$, a strongly negative $\alpha$ is required
if one wants to remove the mass renormalization in this way \cite{WBIH}.
However, then the spectrum moves far away from the GW circle, so we do
not recommend this way to approach the chiral limit.
Positive $\alpha$ increases the pion mass further, but it makes the
shape of the spectrum more circle-like. We are going to take advantage of that.

Next we attach an amplification factor $1/u$ $(u \lsim 1)$ to each link to 
compensate the mean suppression by the gauge field. This is related in spirit to
tadpole improvement \cite{tadpole}, but it can also be viewed as directly
generalizing the tuning of the Wilson hopping parameter.
For our value of $\beta =6$, we reach criticality for the minimally gauged
HF at $u \simeq 0.8$, see Fig.\ \ref{stap-ucrit} (right). Once $u$
is fixed, its inclusion is computationally for free, and it does lead
already to a decent approximation of the GW circle.

\begin{figure}
%\vspace{-3mm}
\begin{tabular}{cc}
\hspace{-8mm}
\def\fpsangle{270}
\epsfxsize=60mm
\fpsbox{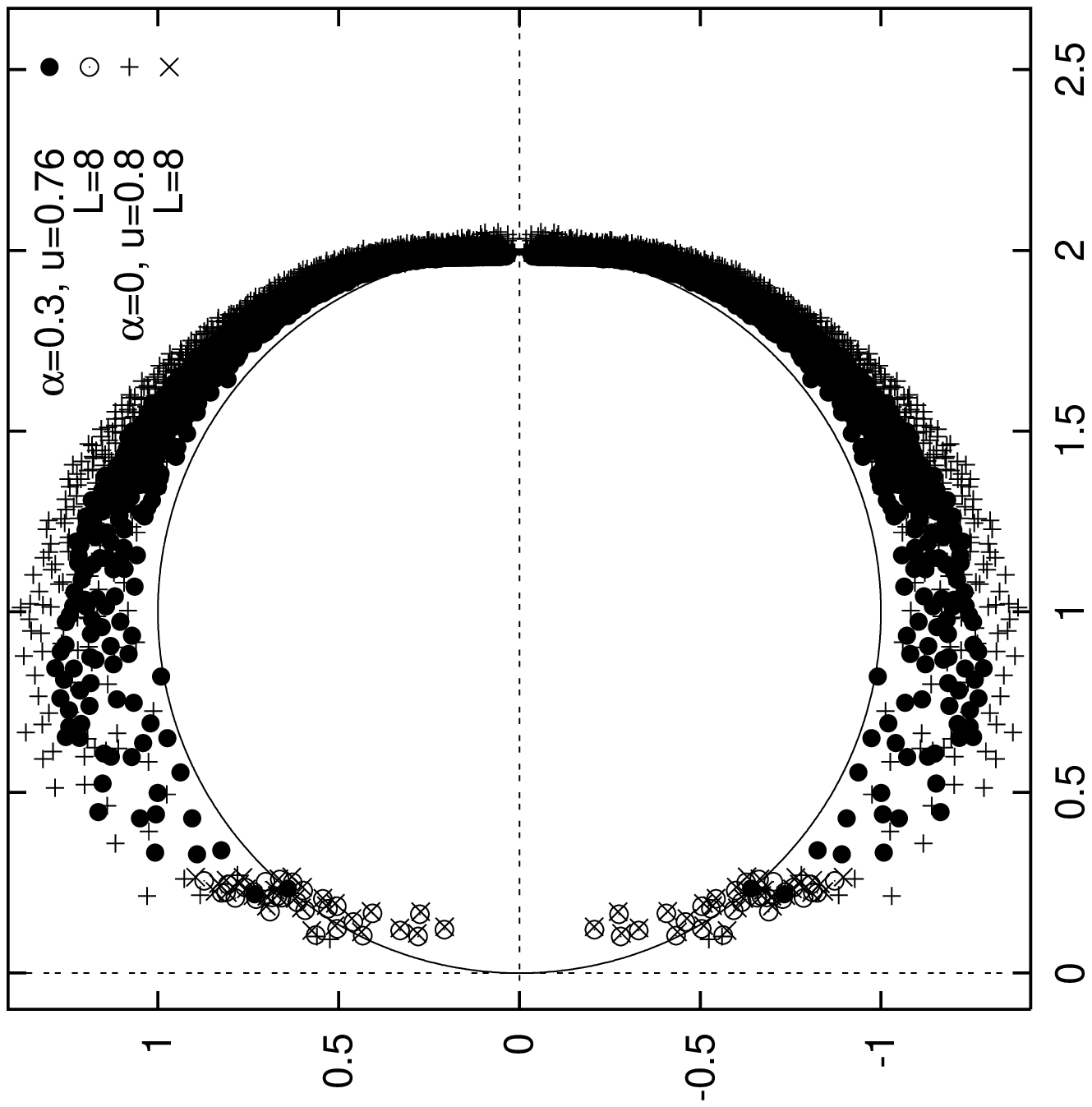}
\hspace{-5mm}
\def\fpsangle{270}
\epsfxsize=60mm
\fpsbox{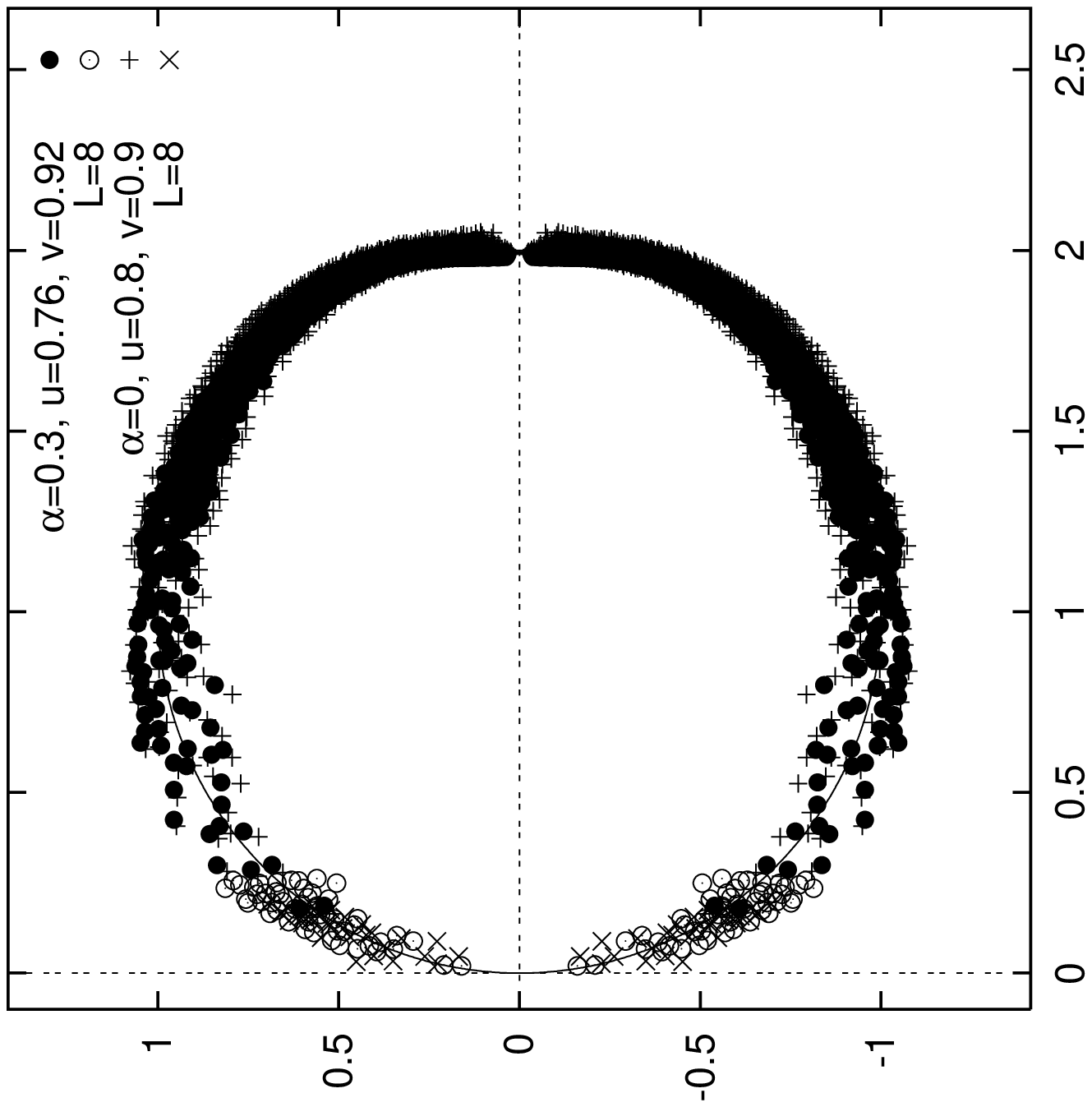}
\end{tabular}
\vspace{-6mm}
\caption{\it{QCD spectra on $4^{4}$ lattices for the HF at critical
link amplification parameter $u$: with and without fat links (left)
and typical results if we further include the kinetic damping parameter $v$ 
at its optimal value (right). We also show the ``continuation'' near 0 in each 
case on an $8^{4}$ lattice with the same parameters, everything at $\beta =6$.}}
\vspace{-5mm}
\label{optimal}
\end{figure}

In a next step, we include fat links with positive staples --- $\alpha =0.3$
is a good value --- and use again the critical link amplification parameter, 
which now amounts to $u \simeq 0.76$. Indeed, this helps to move the eigenvalues
closer to the GW circle, as Fig.\ \ref{optimal} (left) shows. 

Still one would like to further reduce the imaginary part of the
eigenvalues; in particular the upper (and lower) arc still calls for
improvement. Considering the structure of 
$D_{HF}(x,y,U) = \rho_{\mu}(x,y,U)\gamma_{\mu} + \lambda (x,y,U)$, we recognize that
$\rho_{\mu}$ is responsible for the imaginary part, so we multiply a damping factor
$v \lsim 1$ on each link {\em only} in the vector term $\rho_{\mu}$. 
The scalar term $\lambda$, which controls
the left and right arc already successfully, remains untouched.
The optimal value for the new parameter is $v \simeq 0.9$ without fat links,
and $v \simeq 0.92$ at $\alpha =0.3$, always at critical $u$ (which
remains practically unchanged). Now also the upper
arc follows the GW circle closely, and the fat link helps the 
eigenvalues to spread less around the circle, see Fig.\ \ref{optimal} (right).
This is the best approximation achieved so far, and we are confident
that this is about the optimum that can be achieved
with $O(10)$ independent parameters.
In fact, even if we include fat links we are using just 10 independent parameters.

Of course one might still try further parameters, but they should only be included
if they really lead to a significant progress. One could include terms with
a new Dirac structure, and we are currently testing the clover term:
its inclusion (with a positive coefficient)
helps to improve the physically important arc near zero a little,
but it distorts the opposite arc (in the sense that the eigenvalues
fluctuate stronger around the right half-circle).
Generally the clover term tends to attract the eigenvalues closer 
to the real axis, which is also known from the Wilson fermion \cite{GHclov}.
Hence the optimal value of $v$ increases a little: for instance at
clover coefficient 0.15 (and $\alpha =0.3$) it amounts to $v \simeq 0.94$,
while $u \simeq 0.767$ is critical.
%A weak clover term does not change the spectrum much, 
%hence it can be included if this is
%motivated for other purposes (like improved scaling behavior),
%but it does not appear to be a very powerful tool with respect to chirality.

The consideration of the ``magnetic mass'' $m_{B}$ suggests that also a term
$\propto \gamma_{\mu} \gamma_{\nu} \gamma_{\rho}$ could be useful \cite{BBCW}
(we refer to the Pauli term $\vec \sigma \, \vec B / 2m_{B}$ in the low
energy expansion).

What is computationally simple and perhaps promising is an extension
of the fat link to include also (selected) paths of length 5.
All this is currently under investigation and the results will be reported
in Ref.\ [~\raisebox{-6pt}[10pt][6pt]{\Large\cite{prep}}\,].

\section{Summary and Outlook}

Our concept is to first construct a short-ranged approximate
GW fermion with a good scaling behavior and quasi rotational
invariance. Inserting this fermion into the overlap formula makes
the chirality exact. Since the alteration is small, the good scaling
and approximate rotational symmetry is essentially preserved, so that
we obtain an improved overlap fermion. For the same reason, such an overlap
fermion has a high level of locality and its iterative evaluation
is fast. This program was tested extensively in the Schwinger 
model \cite{WBIH}.
Furthermore, for very good approximate GW operators
the interval (for the mass parameter $\mu$)
where the overlap formula can be used safely is extended, 
and such an interval exists up to stronger coupling compared 
to the Neuberger fermion (which inserts the Wilson fermion). 

We add that the same mechanism applies to domain
wall fermions \cite{DWF}: inserting an improved 4d fermion may
induce the same advantages \cite{EPJC,Sha}.

For QCD, we use a truncated perfect free HF as our building block,
and construct a good gauging in the sense of a small GWR violation.
At $\beta =6$ this is achieved, as we observed from
the spectrum: it is close to a GW circle, hence the fermion approximately
obeys the GWR with a point-like kernel $R$.

In our construction, we start form the minimally gauged HF,
remove the additive mass renormalization by a critical amplification
factor for each link, and suppress the vector term only in order to damp
the imaginary part of the eigenvalues, so that the spectrum follows
the shape of a GW unit circle. Fat links help to reduce the fluctuations
of the eigenvalues around that circle, and hence provide an approximation
on a satisfactory level. It seems quite optimal for a set of
about 10 parameters. Perhaps extended staple terms lead to some further
progress, and the clover term helps a little with respect to the fine
resolution close to zero, which we now focus on.
In any case, a good approximation to a GW fermion in QCD has been
accomplished, and it is ready to be inserted into the overlap
formula (\ref{overlap}) \cite{prep}.

\vspace*{-2mm}
\section*{Acknowledgments}
I am very much indebted to Ivan Hip for our ongoing collaboration,
which was the bases of this talk, and for reading the manuscript.
I also thank him, Norbert Eicker and Thomas Lippert
for their work on implementing hypercube fermions in QCD,
and we were glad about the assistance by Christoph Best.
Next I would like to thank the organizers of this workshop, in particular
Xiang-Qian Luo and Eric Gregory. Finally I enjoyed stimulating discussions
related this subject with Ruedi Burkhalter, Philippe de Forcrand,
Yoshio Kikukawa, Martin L\"{u}scher, Gerrit Schierholz and Urs Wenger.

\end{document}